\documentclass[12pt]{article}
\usepackage{jeosman}

\begin{document}

\title{Leaky modes of a left-handed slab}
\author{A. Moreau}
\address{LASMEA, UMR CNRS 6602, Universit\'e Blaise Pascal, 24 avenue
  des Landais, 63177 Aubi\`ere, France.}
\author{D. Felbacq}
\address{GES UMR CNRS 5650, Universit\'e de Montpellier II, Bat. 21,
  CC074, Place E. Bataillon, 34095 Montpellier Cedex 05, France.}
\begin{abstract}
Using complex plane analysis we show that left-handed slab may support either leaky slab waves, which are
backward because of negative refraction, or leaky surface waves, which
are backward or forward depending on the propagation direction of the surface 
wave itself. Moreover, there is a general connection between the reflection
coefficient of the left-handed slab and the one of the corresponding
right-handed slab (with opposite permittivity and permeability) so that
leaky slab modes are excited for the same angle of incidence of the impinging beam for both structures.
Many negative giant lateral shifts can be explained by the excitation of
these leaky modes.
\end{abstract}

\section{Introduction}

Left-handed materials\cite{veselago67} have long been considered a
theoretical oddity. Since it has been demonstrated that they could be
produced using metamaterials\cite{shelby01}, they have attracted much
attention. The basic physics of left-handed materials (LHM) is truly
exotic and has been completely ignored until recently, it renews the
physics of lamellar structures to the extend that a bare slab of
LHM exhibits many surprising properties : 
it can for instance support unusual guided modes\cite{shad03,tichit07} or behave as a
perfect lens\cite{pendry06}. In this paper, we study the exotic
properties of the different types of leaky waves supported by a 
left-handed slab. Given the importance of the left-handed slab for both
fundamental and applied works, there is obviously a need
for a clear understanding of these properties. 

We particularly show that two types of leaky
 waves are supported by such a structure (i) leaky slab waves which 
are always backward {\em due to negative refraction} and 
(ii) leaky surface waves which do not
 exist for a right-handed slab and which can be backward
 or forward. The excitation of these modes leads to positive
or negative giant lateral shifts, the latter being rather 
exotic\cite{tamir}.

\section{Leaky modes and giant lateral shifts}

A leaky mode\cite{tamir} is a solution of the wave equation which verifies the
relation dispersion of a structure but with a propagative solution
above and (or) under the structure. Whereas a guided mode has
a real propagation constant, the propagation constant of a leaky
mode is complex because the energy of the waves leaks out of the
structure and the waves is attenuated. A leaky wave is thus a complex
solution of the dispersion relation and a complex plane analysis
is thus particularly relevant for a thorough analysis of its properties.
Let us underline that a leaky mode may be either forward, which is
common, or backward, leading to a propagation constant which has a
positive (respectively negative) imaginary part.

\begin{figure}[htb]
\centerline{\includegraphics[width=8cm]{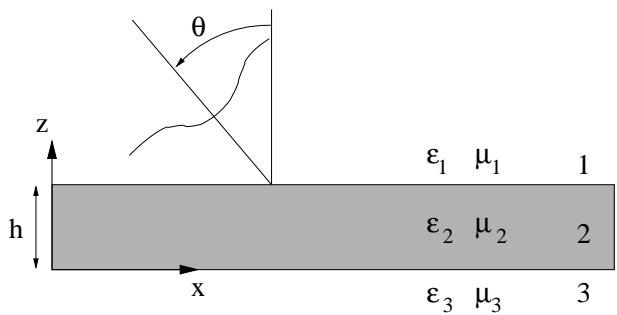}}
\caption{\label{f:schema}The LHM slab of thickness $h$ surrounded by right-handed media.}
\end{figure}

Let us consider a slab characterized by $\varepsilon_2$ and
$\mu_2$ surrounded by right-handed media with $\varepsilon_1$ and $\mu_1$
(resp. $\varepsilon_3$ and $\mu_3$) above (resp. under) the slab as shown figure
\ref{f:schema}. The values we have chosen for $\varepsilon_2$ and
$\mu_2$ are arbitrary but realistic\cite{smith02} so that this structure
could be realized using split-ring resonators and wires.
 
We may assume that $\varepsilon_1\,\mu_1 \geq \varepsilon_3\,\mu_3$ with no
loss of generality. 

The relation dispersion of such a structure can be written 
\begin{equation}\label{e:dispersion}
r_{21}\,r_{23} = \exp(-2i\gamma_2\,h)
\end{equation}
where $\gamma_i=\sqrt{\varepsilon_i\,\mu_i\,k_0^2-\alpha^2}$,
$k_0=\frac{\omega}{c}=\frac{2\pi}{\lambda}$
and $r_{ij} = \frac{\kappa_i-\kappa_j}{\kappa_i+\kappa_j}$ with 
$\kappa_i=\frac{\gamma_i}{\mu_i}$ in TE polarization (or
$\kappa_i=\frac{\gamma_i}{\varepsilon_i}$ in TM polarization).
Since $\varepsilon_1\,\mu_1 \geq \varepsilon_3\,\mu_3$ and we are concerned
with leaky waves, we will only consider values of $\alpha$ such that
$\alpha < \sqrt{\varepsilon_1\,\mu_1}\,k_0$, which means that the
solution will always be propagative at least in medium 1.

Let us now consider the
reflection coefficient of a plane wave $\exp(i(\alpha\,x+\gamma\,z -\omega\,t))$
coming from upwards with an angle of incidence $\theta$
so that $\alpha = n\,k_0 \sin \theta$. Its reflection coefficient can be
written
\begin{equation}
r=\frac{r_{23}\,\exp(2i\gamma_2\,h)-r_{21}}{1-r_{21}\,r_{23}\,\exp(2i\gamma_2\,h)}\label{e:r}
\end{equation}
using the above definitions.

It is obvious that when the relation dispersion is verified, then the
reflection coefficient presents a pole. A leaky mode thus corresponds to
a pole of the reflection coefficient. A zero, located on the other side
of the real axis, corresponds to each pole. As we will see in the
following, a zone where the phase of $r$ quickly varies lies between a pole and its
corresponding zero. This zone crosses the real axis, so that the
presence of a pole is responsible for a swift variation of the phase
on the real axis.

When considering the reflection of a gaussian beam on a structure whose
reflection coefficient has a modulus equal to one (so that it can be
written $r=e^{i\phi}$), the lateral
displacement of the reflected beam's barycenter along the interface is
given by the well known formula 
\begin{equation}\label{e:artmann}
\delta = -\frac{\mbox{d}\phi}{\mbox{d}\alpha}.
\end{equation}
This lateral displacement is the sign that a leaky wave has been excited
by the incident beam. The reflected beam then has two components : the
part which is reflected by the first interface of the structure
(whose barycenter is not particularly displaced) and the leaky wave
itself\cite{tamir}. The reflected beam is heavily distorted by the
leaky wave and presents an exponentially decreasing tail so that
its barycenter is largely displaced : this is the so-called giant lateral shift.

This effect is sometimes called a giant Goos-H\"anchen effect, but in
this case the shift is due to the excitation of a leaky mode\cite{tamir} and not, as in the
real Goos-H\"anchen effect\cite{gh,felbacq03}, to the total reflection.

\section{The left-handed slab}

With left-handed materials, though, negative lateral shifts seem
to be much more common\cite{lakh03,berman02,shad03a,wang05,moreau07} than once
expected\cite{tamir}. Here we will consider the case of a left-handed slab
({\it i.e.} if $\varepsilon_2<0$ and $\mu_2<0$) and explain why the
leaky modes supported by such a structure are usually backward. Our
explanations will be supported by a complex plane analysis of the leaky
modes.

Here the expression (\ref{e:r}) of the reflection coefficient remains
perfectly valid. We will now distinguish two cases : the case when the
solution is propagative in the left-handed medium and the case when the
solution is evanescent in region 2.

\subsection{Leaky slab modes}

\begin{figure}[h]
\centerline{\includegraphics[width=8cm]{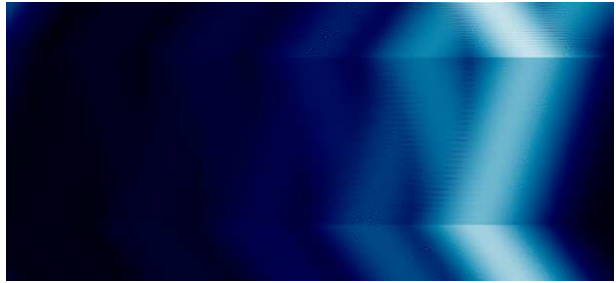}}
\caption{Modulus of the field for a thick left-handed slab with $\epsilon_1=\epsilon_3=\mu_1=\mu_3=1$,
  $\epsilon_2=-3$,$\mu_2=-1$ and $h=60\,\lambda$ using a gaussian incident beam
  with a waist of $20\,\lambda$ and an incidence angle of
  $\theta=45°$.
\label{f:slab_epais}}
\end{figure}

When the field is propagative in the left-handed slab, large negative
lateral shifts have been reported but not interpreted\cite{wang05}.
These shifts are due to the excitation of leaky slab modes or 
Perot-Fabry resonances of the slab at non normal
incidence. Such leaky modes have already been studied for a right-handed
slab\cite{pillon05} and they can be considered as
constructive interferences of the multiple beams which are produced
by reflections on the interfaces of the slab. In the case of a
left-handed slab, since the first beam undergoes a negative refraction
as shown figure \ref{f:slab_epais} these constructive interferences will
logically generate a backward leaky mode. We may thus conclude that
{\it the existence of such a backward leaky mode is linked to negative
refraction}. 

\begin{figure}[htb]
\centerline{\includegraphics[width=8cm]{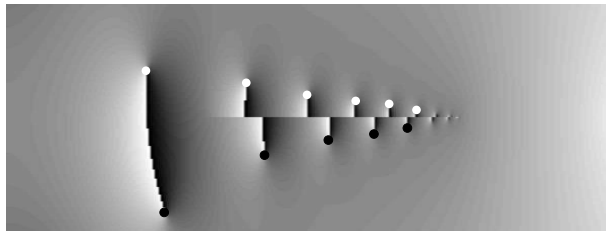}}
\caption{\label{f:phase} The phase of the reflection coefficient in
  a part of the complex plan $[0,n_1\,k_0]+
  i[-\frac{k_0}{\pi},\frac{k_0}{\pi}]$. Each black point represents a pole
  and each white point a zero. The cut line is clearly visible here. The
  rapid variation of the phase which is due to each pole is obvious.}
\end{figure}

This argument is not a proof, though : unexpected lateral shifts have
been reported when the beams interfere destructively\cite{chen04}. But if the
leaky modes are backward, then the corresponding solutions of the
dispersion relation and the poles of the reflection coefficient should
have a negative imaginary part. This is what is shown figure \ref{f:phase}.

Two types of leaky slab waves should be distinguished (i) $L_2$ waves
which are leaky in both the upper and the lower media and (ii) $L_1$ waves
which are leaky only in the upper medium and evanescent in the lower
one. The latter correspond to the poles located under the cut line.

Using complex plane analysis we will now try to show that all the
solutions of the dispersion relation \ref{e:dispersion} are located
in the lower part of the complex plane, meaning that all the leaky modes
are backward.

When the relation dispersion is satisfied, then the following condition
holds :
\begin{equation}\label{e:absr}
|r_{23}\,r_{21}| = e^{2\,\gamma''_2\,h}.
\end{equation}

As demonstrated in the annex, $|r_{ij}>1$ whenever one of the media
is left-handed. Since medium 2 is left-handed then the condition 
\begin{equation}
e^{2\,\gamma''_2\,h} > 1
\end{equation}
should be satisfied, which is possible for $\gamma''_2 > 0$ and therefore
for $\alpha''<0$ (see the annex for details). The
fact that $r_{ij}>1$ is thus the main reason why the poles of $r$ are
under the axis and why the leaky slab modes are backward.

We must underline the fact that our demonstration is valid only for the first
Riemann sheet : our proof cannot exclude that there may be some poles
on the other Riemann sheet above the real axis, corresponding to
forward $L_1$ leaky slab waves when $\varepsilon_1\,\mu_1
>\varepsilon_3\,\mu_3$. But we could not find any.

\subsection{Leaky surface modes}

\begin{figure}[htb]
\centerline{\includegraphics[width=8cm]{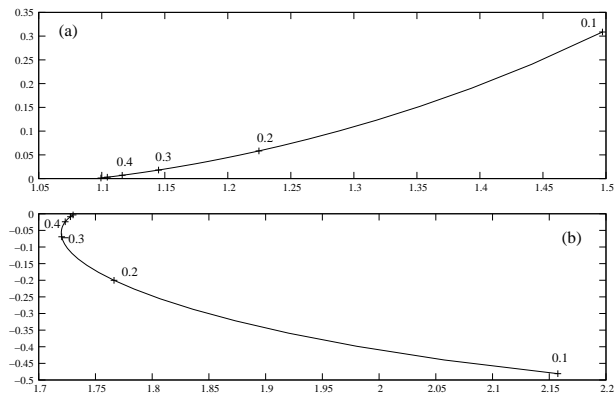}}
\caption{\label{f:poles} 
Location of the poles in the $\frac{\alpha}{k_0}$ complex plane for
 different values of $h$ with 
$\varepsilon_1=9$, $\mu_1=\mu_3=\varepsilon_3=1$
and (a) $\varepsilon_2=-0.5$ and $\mu_2=-1.5$, showing 
a forward surface mode and (b) $\varepsilon_2=-5$ and $\mu_2=-0.5$,
showing a backward surface mode.}
\end{figure}

Let us now consider the situation in which the field is evanescent in the left-
handed medium. Then $\gamma_2$ is purely imaginary on the real
axis. Since $e^{2\,\gamma''_2\,h}$ tends towards infinity when
$h\rightarrow +\infty$ then relation (\ref{e:absr}) can be verified only
if $r_{23}$ has a pole ($r_21$ cannot have one since the field is always
propagative in the upper medium). This means that the structure may
support a leaky mode only if the interface between medium 2 and 3 can
support a guided mode. It is now well-known that such an interface
actually supports a surface mode\cite{ruppin,shad04} which can,
depending on media 2 and 3, be backward (resp. forward)
corresponding to a pole under the real axis (resp. above the real axis
but on the other Riemann sheet). The leaky wave {\em always} has the same
propagation direction as the surface mode, whatever the thickness of
the slab, as shown figure \ref{f:poles}. In the case of a forward leaky wave,
only the zero belongs to the first Riemann sheet, just under the real
axis. The pole shown figure \ref{f:poles} belongs to the other Riemann
sheet.

\begin{figure}[h]
\centerline{\includegraphics[width=8cm]{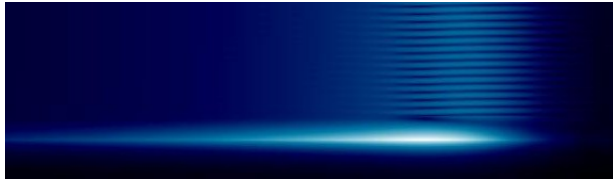}}
\caption{Modulus of the field for a left-handed slab with
  $\epsilon_1=9$,$\epsilon_3=\mu_1=\mu_3=1$,
  $\epsilon_2=-0.5$,$\mu_2=-1.5$ and $h=0.6\,\lambda$ using a gaussian incident beam
  with a waist of $20\,\lambda$ and an incidence angle of
  $\theta=21.496°$. The pole corresponding to the leaky mode is
  located at $\alpha_p=(1.0993+0.001267i)\,k_0$.
\label{f:plasmon}}
\end{figure}

Figure \ref{f:plasmon} finally shows the excitation of a backward leaky 
surface wave by a gaussian beam. The chosen values of $\mu_2$ may be
obtained with simple split ring resonators\cite{soukoulis07} for 
instance.

\section{Fundamental property}

Let us a consider a structure with left-handed materials. We will 
call {\it corresponding right-handed structure} the structure obtained
by replacing any left-handed medium by a medium with
opposite permittivity and permeability, without changing the
geometrical parameters.

In this section, we will concentrate on the link between the
reflection coefficient of a left-handed slab and the one of
its corresponding right-handed structure.

Let us consider the interface between a right-handed medium labelled $i$
and a left-handed medium $j$. The reflection coefficient of such an
interface is $r_{ij}$. We will now define $r^+_{ij}$ the reflection
coefficient of an interface between medium $i$ and right-handed medium
characterized by $|\varepsilon_j|$ and $|mu_j|$. It is not difficult to
see, from the expression of $r_{ij}$ that
\begin{equation}
r^+_{ij} = \frac{1}{r_{ij}}.
\end{equation}

This allows to understand why the Goos-H\"anchen shift of an interface
between a right- and a left-handed medium is the opposite of the
corresponding right-handed structure\cite{berman02} since the phases
of both structures are opposite on the real axis.

The reflection coefficient $r$ can now be written 
\begin{eqnarray}
r&=&\frac{\frac{\,e^{2i\gamma_2\,h}}{r^+_{23}}-\frac{1}{r^+_{21}}}{1-\frac{e^{2i\gamma_2\,h}}{r^+_{21}\,r^+_{23}}}\\
&=& \frac{r^+_{23}\,e^{-2i\gamma_2\,h}-r^+_{21}}{1-r^+_{21}\,r^+_{23}\,e^{-2i\gamma_2\,h}}\\
\end{eqnarray}
Since $\sqrt{z^*} = \sqrt{z}^*$ except when $z$ is on the cut line, then
$\gamma(z^*)=\gamma(z)^*$ and hence $r^+_{ij}(z)^* = r^+_{ij}(z^*)$ so
that
\begin{equation}
r(z)^* = \frac{r^+_{23}(z^*)\,e^{2i\gamma_2(z^*)\,h}-r^+_{21}(z^*)}{1-r^+_{21}(z^*)\,r^+_{23}(z^*)\,e^{2i\gamma_2(z^*)\,h}},
\end{equation}
which can simply be written 
\begin{equation}
r(z)^* = r^+ (z^*),\label{e:fund}
\end{equation}
where $r^+$ is the coefficient reflection of the corresponding
right-handed slab. Note that this relation does not hold on the
cut line, but that it holds for the two Riemann sheets.
This means that the poles of the left-handed slab and the poles of the
corresponding right-handed slab
are symmetrical with respect to the real axis. This
means that $L_2$ waves can be excited for the same incidence angle for both
structures. This is not the case for $L_1$ modes :  the function $r$ on the real axis
is continuous with the lower part of the first Riemann sheet whatever
the situation and the poles which are above the cut line thus have no
effect on the real axis.

\begin{figure}[h]
\centerline{\includegraphics[width=8cm]{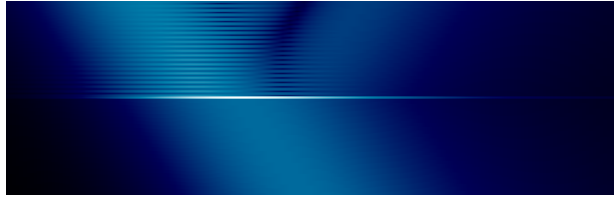}}
\caption{Modulus of the field for a symmetrical slab with
  $\epsilon_1=\epsilon_3=9$, $\mu_1=\mu_3=1$, $\epsilon_2=1.5$,
  $\mu_2=1$ and $h=1.3\,\lambda$ using a gaussian incident beam
with a waist of $20\,\lambda$ and an incidence angle of
  $\theta=22.78°$.
\label{f:l2-forward}}
\end{figure}

\begin{figure}[h]
\centerline{\includegraphics[width=8cm]{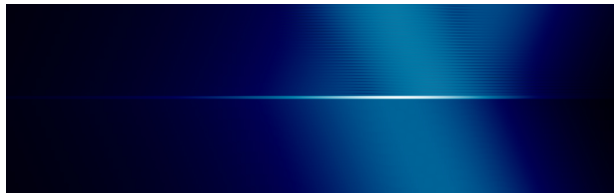}}
\caption{Modulus of the field for a symmetrical slab with
  $\epsilon_1=\epsilon_3=9$, $\mu_1=\mu_3=1$, $\epsilon_2=-1.5$,
  $\mu_2=-1.$ and $h=1.3\,\lambda$ using a gaussian incident beam
with a waist of $20\,\lambda$ and an incidence angle of
  $\theta=22.78°$. The pole corresponding to the leaky mode is
  located at $\alpha_p=(1.16823-0.01125i)\,k_0$       
\label{f:l2}}
\end{figure}

As an example, we have computed the field in TE polarization inside and 
around the slab when it is illuminated with a gaussian beam for the
left-handed slab and its corresponding right-handed structure. The
results are shown figures \ref{f:l2-forward} and \ref{f:l2}.

\section{The grounded left-handed slab}

The grounded left-handed slab is a much more simple structure for (i)
there is no need to distinguish two different types of leaky slab modes
and (ii) the structure can not support any leaky surface mode. All
the leaky modes are then slab modes and are found for $\alpha < n_2\,k_0$. 
The reflection coefficient of the grounded slab is given by (\ref{e:r}) with $r_{23} = -1$ for the TE
polarization and $r_{23} = 1$ for the TM polarization so that the
relation dispersion gives
\begin{equation}
\left| r_{12}  \right| =  e^{2\,\gamma''_2\,h}.
\end{equation}

Since $|r_{12}|>1$ then all the solutions of the dispersion relation are
located in the lower part of the complex plane so that they are all
backward.

It is then easy to show that the relation $r^+(z)^* = r(z^*)$ still holds.
As a consequence, the leaky modes of a grounded left-handed slab and of
its corresponding right-handed structure can be excited for the same
angle of incidence of the impinging beam.

\section{Conclusion}

In this paper, we have thoroughly studied the leaky modes of a
left-handed slab for realistic values of the permittivity and
permeability of the left-handed
medium\cite{smith02,obrien02,soukoulis07} which can be obtained
using structures like split-ring resonators.
Our results can be summarized as follows. Left-handed
slab may support two types of leaky modes :
\begin{itemize}
\item Leaky slab modes, which are always backward because of the negative
      refraction phenomenon. When the transmission is not null,
      leaky modes of the left-handed slab and of its corresponding
      right-handed structure are excited for the same angle of incidence.

\item Leaky surface modes, which may be backward or forward depending on
      the propagation direction of the surface wave itself. 
\end{itemize}

This work could help to interpret many giant lateral
shifts as excitations of exotic leaky waves\cite{wang05,chen04,shad03a}. Since
the existence of backward slab waves is linked to the property of
negative refraction, and since these leaky waves constitute
a signature of a left-handed slab behavior we think that they could
be used to characterize the left-handedness of metamaterial or
photonic crystal structures far better than other methods\cite{kong}.

\section*{Acknowledgments}

This work has been supported by the French National Agency for
Research (ANR), project 030/POEM. The authors would like to thank
Alexandru Cabuz and Kevin Vynck for their help.

\section*{Annex}

In this annex, we will clearly define the choice we have made for
the definition of the complex square root and prove that for $z$ 
on the first Riemann sheet (but not
on the cut line) we have $|r_{ij}(z)|>1$ when media $i$ and $j$ are
not both right-handed.

Since the square root can be continued on the complex plane, $r$ and $r_{ij}$ can
be continued as well. We have chosen to take  
$\sqrt{z} = \sqrt{r}\,e^{i\frac{\theta}{2}}$ with $z=r\,e^{i\theta}$ and $\theta \in
]-\pi,\pi]$, as a definition of the square
root. This means that we have placed the cut line on the negative part of the real axis and if  $x$ is a
positive real, $\sqrt{-x} = i\sqrt{x}$. This
defines the square root on the entire complex plane, to which we refer
as the first Riemann sheet. When we write that $z$ is on the second Riemann
sheet, it will mean that we have taken the opposite of
$\sqrt{z}$ as defined above.

With this choice, we have (i) $\Re(\sqrt{z}) \geq 0$ (ii)
$\sqrt{z^*}=\sqrt{z}^*$ for $z$ on both
sheets but {\em not on the cut line} (iii) if $\Im(z)<0$,
$\Im(\sqrt{z})<0$ and if $\Im(z)>0$, $\Im(\sqrt{z})>0$ (iv) the function 
$\gamma(z)=\sqrt{\epsilon\,\mu\,k_0^2-z^2}$ has a cut line on the real
axis (on $]-\infty,-n\,k_0] \cup [n\,k0,+\infty]$ more precisely) and the function $\gamma$ on the real
axis is continuous with the part of the complex plane which is {\em under
the cut line} : when $z$ passes through the cut line from the
first Riemann sheet (coming from the lower part of the plane) to the
second Riemann sheet, $\gamma(z)$ is continuous. When a function which
can be written using $\gamma(z)$ presents a pole, it must be found
either (i) for $z$ on the first Riemann sheet and under the real axis
(we will say that the pole itself is on the first Riemann sheet in
this case) or (ii) for $z$ on the second Riemann sheet but above the
real axis.

We have 
\begin{equation}
r_{ij}= \frac{\kappa_i-\kappa_j}{\kappa_i+\kappa_j}.
\end{equation}

The modulus of $r_{ij}$ reads as
\begin{eqnarray}
|r_{ij}|^2 &=&
 \frac{(\kappa_i-\kappa_j)\,(\kappa_i^*-\kappa_j^*)}{(\kappa_i+\kappa_j)\,(\kappa_i^*+\kappa_j^*)}\\
&=&\frac{|\kappa_i|^2 + |\kappa_j|^2-2\,(\kappa'_i\,\kappa'_j + 
\kappa''_i\,\kappa''_j)}{|\kappa_i|^2 +
 |\kappa_j|^2+2\,(\kappa'_i\,\kappa'_j + \kappa''_i\,\kappa''_j)},\\
\end{eqnarray}
where $\kappa=\kappa'+i\,\kappa''$.

Let us define $x$ and $y$ the real and imaginary part of $z = x+ i\,y$
on the first Riemann sheet. Let us assume that $x>0$.
We have 
\begin{equation}
\gamma = \sqrt{n^2\,k_0^2-z^2} = \sqrt{n^2\,k_0^2-x^2+y^2 - 2\,i\,x\,y}.
\end{equation}
If $y>0$, then $x\,y >0$ and thus $\Im(n^2\,k_0^2-z^2)<0$ so that finally
$\Im(\gamma)<0$. If $y<0$, then $x\,y<0$ so that $\Im(\gamma)>0$. 
Since $\gamma(-z) = \gamma(z)$ the result will hold
for $x<0$ too and for $x=0$, $\gamma(z)$ is real and positive so that the
result obviously holds. So the
imaginary part of $\gamma(z)$ is positive (resp. negative) when 
the imaginary part of $z$ is negative (resp. positive).

For any right-handed medium, $\kappa$ has the same property than
$\gamma$. For a left-handed medium, since $\kappa=\frac{\gamma}{\mu}$
or $\kappa=\frac{\gamma}{\epsilon}$ depending on the polarization,
the imaginary part of $\kappa$ has the sign of $\Im(z)$.
Since $i$ and $j$ are not both right-handed, then $\kappa''_i$ and
$\kappa''_j$ have not the same sign and the product 
$\kappa''_i\,\kappa''_j$ is always negative. Since 
$\Re(\sqrt{z})>0$ for all $z$ on the first Riemann sheet
then $\kappa'_i\,\kappa'_j$ is always negative too.

Finally, since $\kappa'_i\,\kappa'_j + \kappa''_i\,\kappa''_j<0$, we
have $|r_{ij}|>1$ for all $z$ except on the real axis. Please note that
$r_{ij}$ is not, in the particular case of a left-handed medium, the 
reflection coefficient on the interface\cite{felbacq}.

\bibliographystyle{jeos.bst}
\bibliography{article}

\end{document}